\documentclass[journal=jctcce,manuscript=article]{achemso}

\usepackage{amsmath}
\usepackage{amssymb}
\usepackage{graphicx}
\usepackage{dcolumn}
\usepackage{bm}
\usepackage{color}
\usepackage{booktabs}
\usepackage[T1]{fontenc}
\usepackage{tabularx}
\usepackage{setspace}
\usepackage{listings}
\usepackage{float}
\usepackage{upquote}
\usepackage{soul}

\usepackage{etoolbox}
\BeforeBeginEnvironment{verbatim}{}
\usepackage{siunitx}
\DeclareSIUnit{\molar}{M}

\title[Software for simulations using the CALVADOS model]{Software package for simulations using the coarse-grained  CALVADOS model}


\author{S{\"o}ren von B{\"u}low}   
\affiliation{Structural Biology and NMR Laboratory, Linderstr{\o}m-Lang Centre for Protein Science, Department of Biology, University of Copenhagen, Copenhagen, Denmark}
\email{soren.bulow@bio.ku.dk}

\author{Ikki Yasuda}
\affiliation{Department of Mechanical Engineering, Keio University, Yokohama, Kanagawa, Japan}
\alsoaffiliation{Structural Biology and NMR Laboratory, Linderstr{\o}m-Lang Centre for Protein Science, Department of Biology, University of Copenhagen, Copenhagen, Denmark}
\altaffiliation{Contributed equally to this work, listed in random order}
\author{Fan Cao}
\affiliation{Structural Biology and NMR Laboratory, Linderstr{\o}m-Lang Centre for Protein Science, Department of Biology, University of Copenhagen, Copenhagen, Denmark}
\altaffiliation{Contributed equally to this work, listed in random order}
\author{Thea K. Schulze}
\affiliation{Structural Biology and NMR Laboratory, Linderstr{\o}m-Lang Centre for Protein Science, Department of Biology, University of Copenhagen, Copenhagen, Denmark}
\altaffiliation{Contributed equally to this work, listed in random order}
\author{Anna Ida Trolle}
\affiliation{Structural Biology and NMR Laboratory, Linderstr{\o}m-Lang Centre for Protein Science, Department of Biology, University of Copenhagen, Copenhagen, Denmark}
\altaffiliation{Contributed equally to this work, listed in random order}
\author{Arri{\"e}n Symon Rauh}
\affiliation{Structural Biology and NMR Laboratory, Linderstr{\o}m-Lang Centre for Protein Science, Department of Biology, University of Copenhagen, Copenhagen, Denmark}
\altaffiliation{Contributed equally to this work, listed in random order}
\author{Ramon Crehuet}
\affiliation{Institute for Advanced Chemistry of Catalonia (IQAC), CSIC, Barcelona, Spain}
\altaffiliation{Contributed equally to this work, listed in random order}
\author{Kresten Lindorff-Larsen}   
\affiliation{Structural Biology and NMR Laboratory, Linderstr{\o}m-Lang Centre for Protein Science, Department of Biology, University of Copenhagen, Copenhagen, Denmark}
\email{lindorff@bio.ku.dk}
\author{Giulio Tesei}   
\affiliation{Structural Biology and NMR Laboratory, Linderstr{\o}m-Lang Centre for Protein Science, Department of Biology, University of Copenhagen, Copenhagen, Denmark}
\email{giulio.tesei@bio.ku.dk}

\begin{document}

\begin{abstract}
We present the CALVADOS package for performing simulations of biomolecules using OpenMM and the coarse-grained CALVADOS model. The package makes it easy to run simulations using the family of CALVADOS models of biomolecules including disordered proteins, multi-domain proteins, proteins in crowded environments, and disordered RNA. We briefly describe the CALVADOS force fields and how they were parametrised. We then discuss the design paradigms and architecture of the CALVADOS package, and give examples of how to use it for running and analysing simulations. The simulation package is freely available under a GNU GPL license; therefore, it can easily be extended and we provide some examples of how this might be done.
\end{abstract}

% \maketitle

\section{Introduction}

\subsection{Coarse-grained molecular models for simulations of disordered and multi-domain biomolecules}

Intrinsically disordered proteins and regions in proteins (IDPs and IDRs) are important for biological function and involved in various diseases \citep{holehouse2024molecular}. 
Around 30\% of the residues in the human proteome are predicted to be disordered with around 70\% of human proteins containing at least one long (>30 residues) IDR, and some proteins are fully disordered in vitro and in the cell \citep{holehouse2024molecular,pritisanac2024functional, moses2024structural}.
IDRs adopt broad sets of interconverting configurations, and the properties of such conformational ensembles are modulated by the amino acid sequence and solution conditions, and influence phase behaviour and protein function in the cell \citep{holehouse2024molecular}.

Molecular dynamics (MD) simulations can be used to examine the behaviour of biomolecules at high spatial and temporal resolution. MD simulations numerically integrate a set of equations of motion using interaction potentials (force fields) that are typically parametrised using a combination of experimental data and higher-level (for example quantum-chemical) calculations. Historically, disordered proteins are difficult targets for atomistic molecular simulations. This difficulty arises from two main challenges: The force field problem and the sampling problem. Force fields for atomistic biomolecular simulations were originally mostly tested to model short peptides or the folded states of proteins, and were later found to give rise to unphysically compact IDR ensembles and overly attractive protein-protein interactions \citep{petrov2014are}. Several more modern force fields can describe both folded and disordered proteins relatively well \citep{huang2017charmm36m,robustelli2018developing,piana2020development}. However, some inaccuracies persist, including in capturing the global chain dimensions, which vary significantly with the choice of water model \citep{sarthak2023benchmarking}. The sampling problem relates to the challenge of exhaustively sampling protein conformational states. For a simulation of a protein in explicit solvent, most computational resources are spent on calculating water interactions. This is exacerbated by the large simulation boxes needed to minimise finite-size effects in simulations of disordered molecules with extended conformations that might interact across periodic images.

One approach to enhance sampling of conformational space is to simplify the description of the protein, water, or both. Below we describe some of these models that have been applied to IDPs, but do not intend to provide a comprehensive overview of the many models that are available. In models such as ABSINTH \citep{vitalis2009absinth}, PROFASI \citep{irback2009effective} and CHARMM EEF1 \citep{lazaridis1999effective,bottaro2013variational} the protein is described in atomistic detail, but with a continuum model for the solvent. In models such as Martini \cite {souza2021martini}, SPICA \citep{kawamoto2022spica} and SIRAH \citep{darre2015sirah} both protein and water molecules are described with a coarse-grained representation, and updated versions have been described that better capture larger-scale conformational properties of IDPs \citep{florencia2021sirah,thomasen2022improving,thomasen2024rescaling,yamada2023improved, wang2025martini3idp}. Despite the reduced number of particles and allowing for larger time steps in MD, these force fields can still be computationally demanding, both for large systems and for single-chain simulations of long IDPs.
%Not sure if it is worth mentioning AWSEM-IDPs, https://pubs.acs.org/doi/10.1021/acs.jpcb.8b05791 with a similar resolution as Martini but having implicit solvent

A set of models coarse-grain even further by combining a coarse-grained model of the protein with a continuum representation of the solvent. Several of these represent the protein by two or more beads per residue and have been applied to study IDPs \citep{wu2018awsem,zhang2023toward,mugnai2025sizes}. Here we instead focus on models where residues are mapped onto single beads \citep{ashbaugh2008natively,norgaard2008experimental, kim2008coarse,dignon2018sequence,latham2019maximum, tesei2021accurate,dannenhoffer2021data, joseph2021physicsdriven, jussupow_cocomo2_2025}; we note that several other such models exist. We here refer to these as hydrophobicity scale (HPS) models, although the term HPS was originally meant to indicate a specific set of parameters in one such model \citep{dignon2018sequence}. In HPS models, the solvent is treated implicitly as a dielectric continuum. To account for water-mediated interactions, the pairwise potentials between residues are scaled according to their hydropathy or `stickiness', as defined by a hydrophobicity scale. Various HPS models have been developed, differing in the hydrophobicity scale used, the treatment of charges \citep{tejedor2025chemically}, and the use of additional interaction terms like dihedral angle potentials \citep{rizuan2022developing}.

Simulations with HPS models are computationally efficient enough to be applied to study conformational properties of thousands of isolated IDRs at the proteome scale \citep{lotthammer2024direct, tesei2024conformational}, to perform hundreds of simulations one after another \citep{pesce2024design}, and to simulate hundreds of chains to predict the propensity of proteins to undergo phase separation \citep{dignon2018sequence,regy2021improved, tan2023highly,  tesei2023improved, an2024active}. As such, simulations with HPS models have also been used to train or benchmark models that predict biophysical properties of IDRs from their sequence \citep{zheng2020hydropathy,tesei2024conformational,lotthammer2024direct,houston2024physics, von2025prediction} and deep learning models that generate conformational ensembles directly from sequence \citep{lewis2024scalable, janson2024transferable, zhu2024precise, zhang2025deep,novak2025accurate}.

HPS models were originally designed and tested for disordered proteins, but have also been applied or tested more extensively to study multi-domain proteins (MDPs), consisting of folded domains connected by flexible linkers \citep{dignon2018sequence,krainer2021reentrant, cao2024coarsegrained,jussupow_cocomo2_2025}, and to disordered single-stranded RNA \citep{regy2020sequence, joseph2021physicsdriven, valdes-garcia2023modeling, yasuda2024coarsegrained}.

\subsection{The CALVADOS force fields}
We have developed a set of CALVADOS (Coarse-graining Approach to Liquid-liquid phase separation Via an Automated Data-driven Optimisation Scheme) models for simulations of proteins and other molecules. Here, we briefly describe the general pair potentials of this family of HPS models, which include the CALVADOS~2 force field for IDRs \citep{tesei2021accurate, tesei2023improved} and its extensions to MDPs (CALVADOS~3) \citep{cao2024coarsegrained}, RNA \citep{yasuda2024coarsegrained}, and PEG \citep{rauh2025crowder}. Further details on molecule-specific interactions are covered in the next section.

% The force fields in the CALVADOS family are HPS models with a one-bead-per-residue representation for proteins and a two-bead-per-residue representation for RNA \citep{tesei2021accurate, tesei2023improved, cao2024coarsegrained, yasuda2024coarsegrained}. All molecules have bonded and nonbonded interaction terms. Protein and RNA share the same functional forms for bonds and for nonbonded interactions, whereas RNA has an additional angle potential and a stacking potential between neighbouring bases \citep{yasuda2024coarsegrained}.

Bonds between residues are described by a harmonic potential,
\begin{equation}
\label{eq:2}
u_\text{bond}(r)=\frac{1}{2}k(r-r_0)^2,
\end{equation}
where $k=\SI{8033}{\kJ \mol^{-1} \nano\meter^{-2}}$ is the force constant and $r_0$ is the molecule-specific equilibrium bond distance.

Nonbonded non-ionic interactions are described by the Ashbaugh-Hatch (AH) potential \cite{ashbaugh2008natively}, a modified Lennard-Jones (LJ) potential that effectively accounts for any non-ionic interaction, such as hydrophobic interactions, $\pi$-$\pi$ stacking, and hydrogen bonding. Key parameters of this potential are the amino acid-specific diameters, $\sigma$, and stickiness values, $\lambda$, which together influence the strength and the range of the interaction. In CALVADOS, the AH potential is truncated and shifted at $r_{c,\text{AH}}=2$ nm \citep{tesei2023improved},
\begin{equation}
\label{eq:5}
u_\text{AH}(r)=\left\{
\begin{aligned}
&u_\text{LJ}(r)-{\lambda}u_\text{LJ}(r_{c,\text{AH}})+{\epsilon}(1-\lambda),& r\leq2^{1/6}\sigma \\
&{\lambda}[u_\text{LJ}(r)-u_\text{LJ}(r_{c,\text{AH}})],                    & 2^{1/6}\sigma<r\leq r_{c,\text{AH}}\\
&0,                                                   & r>r_{c,\text{AH}}
\end{aligned}
\right.
\end{equation}
where $\sigma = (\sigma_i + \sigma_j)/2$, $\lambda = (\lambda_i + \lambda_j) / 2$ for residues $i$ and $j$, and the classic LJ potential, 
\begin{equation}
\label{eq:6}
u_\text{LJ}(r)=4{\epsilon}\left[\left(\frac{\sigma}{r}\right)^{12}-\left(\frac{\sigma}{r}\right)^6\right],
\end{equation}
with ${\epsilon}=\SI{0.8368}{\kJ\per\mol}$.

The $\lambda$ parameters are key ingredients in the CALVADOS force field, as they capture the effective interactions between amino acid residues. We have developed an approach to learn force field parameters from experimental data \citep{norgaard2008experimental} and used similar procedures to learn the $\lambda$ values in the CALVADOS protein force fields \citep{tesei2021accurate,tesei2023improved,cao2024coarsegrained}.

Solvent-mediated salt-screened charge-charge (ionic) interactions are modelled via the Debye-H{\"u}ckel (DH) potential, truncated and shifted at $r_{c,\text{DH}}=4$ nm,
\begin{equation}
\label{eq:3}
u_\text{DH}(r)=\left\{
\begin{aligned}
&\frac{Z_iZ_j e^2}{4\pi\epsilon_0\epsilon_r}\left[ \frac{\exp(-r/D)}{r} - \frac{\exp(-r_{c,\text{DH}}/D)}{r_{c,\text{DH}}} \right ],& r\leq r_{c,\text{DH}} \\
&0,                                                   & r>r_{c,\text{DH}},
\end{aligned}
\right.
\end{equation}
Here, $e$ is the elementary charge, $Z_i$ and $Z_j$ are the charge numbers of beads $i$ and $j$, $\epsilon_0$ is the vacuum permittivity, $D=\sqrt{1/(8{\pi}BI)}$ is the Debye length of an electrolyte solution of ionic strength $I$, and $B({\epsilon}_r)$ is the Bjerrum length of the temperature-dependent dielectric constant $\epsilon_r$ \cite{akerlof1950dielectric},
\begin{equation}
\label{eq:4}
\begin{split}
{\epsilon}_r(T)=\frac{5321}{T}&+233.76-0.9297\,T \\ 
&+1.417{\times}10^{-3}\,T^2-8.292{\times}10^{-7}\,T^3.
\end{split}
\end{equation}
To model the effect of different solution pH values, we set the charge of the histidine residues using the Henderson-Hasselbalch equation,
\begin{equation}
Z_\text{His}= \frac{1}{1+10^{\text{pH}-\text{p}K_{a}}},
\end{equation}
with $\text{p}K_{a}=6.00$.

\subsection{Additional molecule-specific CALVADOS parametrization}

The details of the models describing folded protein domains, disordered RNA and polyethylene glycol (PEG) crowding have been described in detail before \citep{cao2024coarsegrained, yasuda2024coarsegrained, rauh2025crowder} and will only be summarized here. 

Briefly, folded domains are manually restrained using a harmonic potential between nonbonded pairs of residues within a cutoff of \SI{0.9}{\nano\meter}. The equilibrium distance of such a restraint is set to the centre-of-mass (COM) separation calculated from a structure that is used as input. For MDPs consisting of folded domains connected by IDRs, we showed that the COM representation reduces overly attractive domain--domain interactions and thereby prevents the compact ensembles observed for some proteins when using the C$_\alpha$ representation \citep{cao2024coarsegrained}. Therefore, we use a mixed representation for MDPs, where residues in folded domains are represented by their COMs and those in IDRs by their C$_\alpha$ atoms, using the same C$_\alpha$-C$_\alpha$ equilibrium distance of \SI{0.38}{\nano\meter} as for the CALVADOS~2 force field. Using this mapping, we reoptimised the $\lambda$ parameters of the model and obtained CALVADOS~3, which performs on par with CALVADOS~2 for IDPs while improving the model accuracy for MDPs.

Synthetic crowders are often used to probe the effect of nonspecific macromolecular crowding on the dynamics of proteins, including their phase-separation propensity. We have developed a model for PEG to study the effect of crowding on conformational and phase properties of IDRs \citep{rauh2025crowder}. The size and stickiness of the individual PEG residues (`monomers') were optimised against experimental data reporting on the single-chain compaction of isolated PEG and of IDRs at varying concentrations of PEG. The model can, for example, be used to perform simulations of the phase behaviour of protein systems that do not easily form condensates in the absence of crowding agents.

Disordered RNA is modelled in CALVADOS with a two-bead-per-residue representation to separate the effects of the non-sticky negatively charged backbone and aromatic nucleobases \citep{yasuda2024coarsegrained}. CALVADOS-RNA was parametrised using a combined bottom-up and top-down approach against atomistic MD simulations and experimental data, respectively. In addition to the standard pair potentials used for the protein model, CALVADOS-RNA includes a stacking term between neighbouring nucleobases and an angle potential to reproduce local geometry distributions from atomistic simulations \citep{bergonzo2022conformational}. The AH parameters for backbone and nucleobases were subsequently optimised to match experimental radii of gyration, $R_g$, and second virial coefficients, $B_2$ \citep{yasuda2024coarsegrained}. The RNA model was specifically tested to be compatible with the CALVADOS~2 protein force field, enabling simulations of condensates formed by RNA--protein mixtures.

\section{Architecture of the CALVADOS package}

\subsection{General design}

The CALVADOS package is designed to streamline and simplify the process of setting up, simulating, and analysing coarse-grained systems of varying complexity using the CALVADOS models described above. As a minimum example, only the sequence, number and type of molecules, simulation box dimensions, and solution conditions need to be supplied by a user to run a simulation. Conversely, the package enables advanced users to set up complex systems and to define new types of molecules or residues (e.g., post-translational modifications or cyclic peptides).

\begin{figure}[tb]
    \includegraphics[width=4.5in]{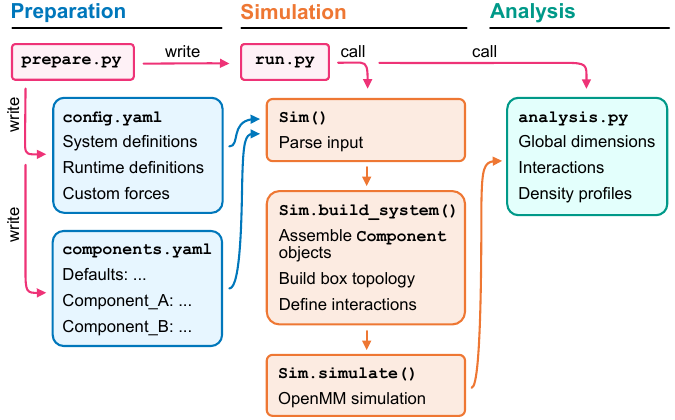}
    \caption{Architecture of the CALVADOS package. A wrapper script \texttt{prepare.py} generates the files \texttt{config.yaml}, \texttt{components.yaml}, and \texttt{run.py} for general definitions, molecule definitions, and running the simulation, respectively. \texttt{run.py} calls the main class \texttt{Sim}, which parses the input, handles setting up the system, and runs the OpenMM simulation. \texttt{run.py} can optionally be configured to call analysis scripts after the simulation is completed. Various trajectory post-processing and analysis routines for single-chain and multi-chain properties are included in the package.}
    \label{fig:1}
\end{figure}

We chose the OpenMM \citep{eastman2017openmm} simulation software as the backend, both for its flexibility and for its Python API. We do not make use of the xml-based force field description implemented in OpenMM. Instead, to ease the entry barriers for new users, the CALVADOS package automatically parses sequence input into an OpenMM-readable system topology. Fig.~\ref{fig:1} shows the overall architecture of the software. The main modules of the package are \verb|sim.py| and \verb|components.py|, which deal with the overall setup of the simulation system and the definitions of the molecules (see Sections~\ref{sec:sim_class} and \ref{sec:component_class}). Additional modules include functions related to input parsing, interaction potentials, sequence parsing and manipulation, building of molecular configurations, as well as postprocessing and analyses.

\subsection{User input}

The user generally provides two input files: A system configuration file (default: \verb|config.yaml|) and a component file (default: \verb|components.yaml|). The system configuration file describes global parameters such as box dimensions, temperature, and pH. The component file defines the types and numbers of molecules together with other molecule-specific properties such as whether to account for charged moieties at the termini of polypeptide chains. A default section in the \verb|components.yaml| configuration file can be used to define properties shared by all molecules. Molecule-specific definitions overwrite the default, allowing users to mix default settings with molecule-specific input. Additional input files such as custom residue definitions, folded domain definitions or custom restraints can be required depending on the specifications of the system and/or components.

The \verb|config.yaml| and \verb|components.yaml| input files can be written and edited manually. However, our package provides the Python wrapper \verb|prepare.py| that conveniently generates both files together with a \verb|run.py| script to start the simulation, and, optionally, batch job submission files to run the simulation on a server. The wrapper can be supplied with minimal definitions which are then combined with the default settings specified in the files \verb|calvados/data/default_config.yaml| and \verb|calvados/data/default_component.yaml|. Template job configuration files are also available (\verb|calvados/data/default_job.yaml|) alongside batch submission script templates in \verb|jinja| format (\verb|calvados/data/templates/|). New job templates can be added to match the user's server architecture. The \verb|cfg.py| script manages the settings in the wrapper and writes the input files. The examples in Section~\ref{sec:examples} show a minimum wrapper (Section~\ref{sec:ex_single_IDR}) and extensions thereof for different systems.

\subsection{\texttt{Sim} class} \label{sec:sim_class}

% The code architecture of CALVADOS mirrors the two main input files. 
The system setup is split into the \verb|Sim| class of the \verb|sim.py| module and the \verb|Component| class of the \verb|components.py| module. All general definitions and definitions pertaining to multiple components (e.g., intermolecular interactions) are dealt with by \verb|Sim|, whereas all definitions at the level of a single molecule (e.g., bond connectivity, bead properties, geometry of the initial configuration, intramolecular restraints, etc.) are defined in the \verb|Component| class.

The \verb|Sim| class parses the input and sets up the simulation system. First, \verb|Sim.__init__()| reads and processes the system configurations (\verb|config.yaml|) and component (\verb|component.yaml|) input files. \verb|Sim| requires all relevant simulation parameters to be defined in these files.

Following input parsing, \verb|Sim.build_system()| defines simulation box dimensions and periodic boundary conditions. For each \textit{different} molecular component of the system, \verb|Sim| then instantiates an object of the \verb|Component| class (and subclasses thereof) in the \verb|components.py| module, described in Section~\ref{sec:component_class}.

Molecules are then distributed in the box based on the keyword \verb|topol| to generate a starting configuration (e.g., with molecules placed in the centre, randomly, in a slab, or on a grid), which is saved as a structure file in PDB format. Non-bonded interactions are defined and the general definitions, including information from each \verb|Component| object, are packaged into an \verb|openmm.System| object.

The \verb|simulate()| method of \verb|Sim| combines the input configuration, \verb|openmm.System| object, a Langevin integrator (by default with time step $t=\SI{10}{\femto\second}$ and friction coefficient $\gamma=\SI{0.01}{\pico\second}^{-1}$), and a desired platform (CPU, OpenCL, or CUDA) to create an \verb|openmm.Simulation| object. Finally, the simulation is run for a set number of steps or simulation time in hours.

\subsection{\texttt{Component} class} \label{sec:component_class}

In CALVADOS, a \verb|component| refers to a specific molecule (e.g., the protein FUS) with attributes including sequence, number of molecules, charges, geometry, connectivity, and bond forces. Every \textit{different} molecule has its own \verb|component| object, whereas multiple copies of the exact same molecule belong to the same \verb|component| object. For example, four FUS proteins share the same \verb|component| object, whereas an additional single Ddx4 protein would have a separate \verb|component| object. 

Each type of biomolecule (protein, RNA, etc.) has its own subclass (\verb|Protein|, \verb|RNA|, etc.) which inherits from the \verb|Component| class. The design principle behind this is that certain attributes between the different biomolecules are shared. For example, all the molecule types have a sequence, particle beads, a molecular weight, etc., whereas other properties are molecule-type specific.

A generic molecule in \verb|Component| is processed as follows: Read in the sequence, calculate the number of residues and number of beads, determine molecular weights, determine particle bead sizes and $\lambda$ stickiness parameters for use in the AH potential, determine charges for use in the DH potential, and determine bond lengths. Finally, a geometry of the molecule is either read from the PDB file or built from scratch.\footnote{For example, IDRs are by default packaged in a `compact' representation resembling a cube. Since configurations relax quickly during the minimisation and simulation runs, the starting geometry is typically not important and is optimised to package as many molecules as possible into the simulation box without clashes.}

The subclasses, such as \verb|Protein| or \verb|RNA|, incorporate the definitions that are biomolecule specific by adding additional functions or overwriting default functions of \verb|Component|, where needed. For example, the number of beads per residue and the connectivity of beads (define `what a bond is') differ between the one-bead-per-residue protein model and two-bead-per-residue RNA model, as does the treatment of terminal residues in the \verb|RNA| subclass. \verb|RNA| also introduces additional angle and stacking potentials. In contrast, the \verb|Protein| subclass has additional routines for restraining folded domains. This modularity allows for easy modification and addition of molecule definitions (see Section~\ref{sec:extension}).

\section{Tutorial examples} \label{sec:examples}

Below, we provide a number of examples to illustrate some of the types of applications that are made possible with the CALVADOS package. We note that our goal is generally not to motivate the science extensively or to discuss the results, but rather to illustrate how the package can be used and extended. We also stress the importance of reading the original literature for more details on the approximations involved, the ranges of applicability, and the extent to which different applications have been validated. All example scripts described in this section can be run from the \verb|examples| folders on \verb|github.com/KULL-Centre/CALVADOS|.

\subsection{Single-chain IDR simulation} \label{sec:ex_single_IDR}
Using CALVADOS, we can simulate single IDRs and accurately predict their conformational properties providing as input the amino acid sequence, temperature, ionic strength, and pH of the buffer solution. In this example, we simulate the low-complexity domain (LCD) of the heterogeneous nuclear ribonucleoprotein (hnRNPA1$^*$), hereafter A1-LCD$^*$, with * indicating a sequence missing the hexa-peptide 259--264. hnRNPA1$^*$, which is a splicing factor, consists of two RNA recognition motifs connected by a short linker, followed by the LCD: a 131-residue disordered C-terminal domain. This architecture is characteristic of many ribonucleoproteins which, in response to cell stress, together with RNA may form biomolecular condensates known as stress granules \cite{molliex2015phase}. A1-LCD$^*$ has been shown to be necessary and sufficient for the phase separation of hnRNPA1$^*$ in vitro \cite{molliex2015phase,martin2021interplay}, and mutations within this region are implicated in neurodegenerative diseases, such as amyotrophic lateral sclerosis \cite{kim2013mutations}. The sequence features determining the compaction and phase properties of A1-LCD$^*$ have been studied extensively \cite{martin2020valence,bremer2022deciphering}.

To simulate a single copy of A1-LCD$^*$, we initialise the \verb|Config| class in the \verb|prepare.py| script as follows:
\begin{verbatim}
config = Config(
    box=[50, 50, 50],  # nm
    topol='center',
    temp=293,          # K
    ionic=0.19,        # M
    pH=7.5, 
    wfreq=7000,        # trajectory writing interval
                       # 1 step = 10 fs
    steps=1010*7000,   # 1010 frames
    )
\end{verbatim}
With the options \verb|box| and \verb|topol|, we place the IDR in the centre of a cubic box with 50-nm side length and specify \verb|temp|, \verb|ionic|, and \verb|pH| to set the same temperature, ionic strength, and pH as in the experimental SAXS measurements by Martin et al. \cite{martin2021interplay}. By setting \verb|ionic| equal to 0.190~M, we account for ionic strength contributions from salt (150~mM NaCl) as well as from the buffer (50~mM Tris at pH~7.5 contributes with $\approx40$~mM to the ionic strength). We save frames to a trajectory \verb|DCD| file every 7,000 MD steps, corresponding to $\Delta t=70$ ps. We run the simulation for 70,700~ps and discard the first 700~ps. These settings allow us to collect 1,000 weakly correlated conformations. In a previous study, we fine-tuned the saving frequency to sample consecutive configurations with a low extent of self-correlation irrespective of sequence length, $N$, and obtained $\Delta t \approx 3 \times N^2\,\si{\femto\second}$ if $N>150$ and $\Delta t=70\,\si{\pico\second}$ otherwise \cite{tesei2024conformational}. With this empirical relationship, we found that the value of the autocorrelation function of the radius of gyration, for a time lag of one frame, plateaus to $\sim0.5$ for $N>200$. 

In the \verb|prepare.py| script, we initialise default \verb|Components| definitions as follows:
\begin{verbatim}
components = Components(
    molecule_type='protein',
    nmol=1,                                   
    charge_termini='both',                    
    fresidues=f'{cwd}/input/residues_CALVADOS2.csv', 
    ffasta=f'{cwd}/input/idr.fasta',                 
    )
\end{verbatim}
\verb|molecule_type| and \verb|nmol| specify that we are simulating a single copy of a protein. With \verb|charge_termini='both'|, we add and subtract a unit charge to the N- and C-terminus, respectively, accounting for the extra positive charge and negative charge on the ammonium group at the N-terminus and the carboxylate at the C-terminus. N-, C-, and end-capped proteins can instead be simulated with the options \verb|charge_termini='C'| (only adding a charge to C), \verb|'N'| (only adding a charge to N), and \verb|'end-capped'| (not adding any charge), respectively. 
We then specify the paths of the file containing the amino acid-specific parameters with \verb|fresidues|, here parametrised using the CALVADOS~2 force field \citep{tesei2023improved}, and the FASTA file with the IDR sequence with \verb|ffasta|, where \verb|{cwd}| should point to the current working directory.\footnote{Note that all path definitions are relative to the folder that the simulations are started from. To ensure that input files are found regardless of where the simulation is started, we always recommend creating absolute pointers to all input files using \texttt{import os; cwd = os.getcwd();} and \texttt{f'\{cwd\}/...'} in the \texttt{prepare.py} script. The provided example files follow this behaviour.}

The IDR of choice (A1SLCD in this case) is then added with
\begin{verbatim}
components.add(name='A1SLCD')
\end{verbatim}
using the default settings defined above. All component defaults can be overwritten during the \verb|components.add()| statement. This can be useful for multiple different components with, e.g., different numbers of molecules but otherwise equal definitions.

\begin{figure}[tb]
\centering
\includegraphics[width=4.5in]{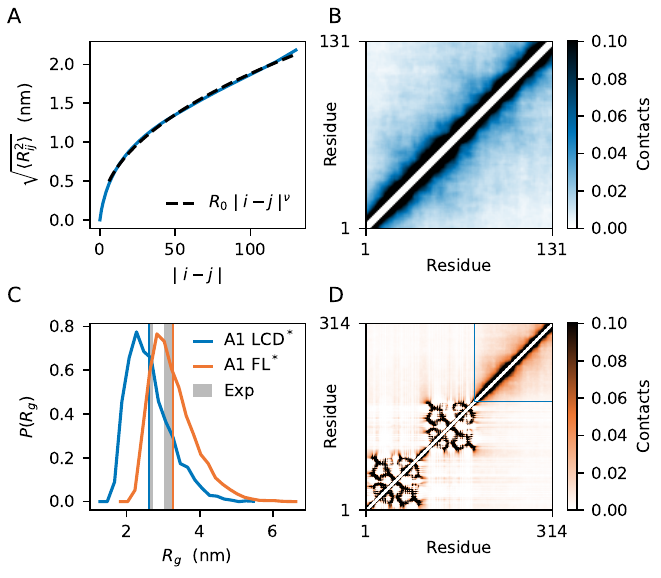}
\caption{Single-chain simulations. (A) Root-mean-square residue-residue distances, $\sqrt{\langle R_{ij}^2 \rangle}$, vs. separations along the linear sequence, $|i-j|$, calculated from single-chain simulations of A1-LCD$^*$. Dashed line: nonlinear fit to $\sqrt{\langle R_{ij}^2 \rangle}=R_0|i-j|^\nu$. (B) Intra-chain contact map of A1-LCD$^*$. (C) Distributions of the radius of gyration, $R_g$, of A1-LCD$^*$ and full-length hnRNPA1$^*$. Vertical solid lines: ensemble-averaged simulated $R_g$. Vertical grey bars: confidence interval of the experimental $R_g$ values measured by Martin et al. \cite{martin2021interplay}. (D) Intra-chain contact map of full-length hnRNPA1$^*$. Blue box: LCD from residue 183 to residue 314. All simulation data shown in this figure are averaged over three independent replicas.}
\label{fig:2}
\end{figure}

After running the simulations, the CALVADOS package also helps analyse the simulation trajectories, for example to calculate the internal-distance scaling exponent, $\nu$, the intra-chain residue-residue contacts, the end-to-end distance, $R_{ee}$, and the radius of gyration, $R_g$. We can perform these analyses right after the simulation run by adding the following lines in \verb|prepare.py|:
\begin{verbatim}
subprocess.run(f'mkdir -p A1SLCD', shell=True)
subprocess.run(f'mkdir -p data', shell=True)
analyses = f"""
from calvados.analysis import save_conf_prop
save_conf_prop(
    path='A1SLCD', name='A1SLCD',
    residues_file=f'{cwd}/input/residues_CALVADOS2.csv',
    output_prefix='data',
    start=10, is_idr=True, select='all'
    )
"""
config.write(path, name='config.yaml', analyses=analyses)
\end{verbatim}
First, we create directories to store the trajectory (\verb|A1SLCD|) and the analysis files (\verb|data|). Second, we write the script \verb|run.py| appending to it two lines that, once the simulation has completed, import and call the function \verb|save_conf_prop()|, which calculates per-frame $R_{ee}$ and $R_g$ values, the average contact map, and the average $\nu$ (Fig.~\ref{fig:2}A--C). We estimate $\nu$ from a nonlinear fit to $\sqrt{\langle R_{ij}^2 \rangle}=R_0|i-j|^\nu$ of the root-mean-square residue-residue distances, $\sqrt{\langle R_{ij}^2 \rangle}$, for separations along the linear sequence, $|i-j|$, larger than 5 (Fig.~\ref{fig:2}A). To obtain the contact map (Fig.~\ref{fig:2}B), we calculate residue-residue distances, $r_{ij}$, for $|i-j|>3$ and apply the switching function
\begin{equation} \label{eq:switch}
    c(r_{ij}) = 0.5-0.5\tanh{[(r_{ij}-r_0)/w]}
\end{equation}
where $r_{ij}$ is the intermolecular distance between two residues, $r_0=\SI{1}{\nano\meter}$, and $w=\SI{0.3}{\nano\meter}$. The CALVADOS package uses MDAnalysis \citep{michaud2011mdanalysis} and MDTraj \citep{McGibbon2015MDTraj} to help in analysing the trajectories, and the user can use these and many other tools to analyse the trajectories in other ways.

A simulation performed with the settings detailed in this section takes 25 min on a single core of an AMD EPYC 9754 CPU and 3 min on an NVIDIA A40 GPU.

We performed three independent simulations of A1-LCD$^*$ and estimated ensemble averages and the corresponding errors as the mean $\pm$ the standard deviation (SD) across the replicas. From the simulations we calculate an apparent scaling exponent of $\nu=0.473\pm 0.001$ (Fig.~\ref{fig:2}A), indicative of a compact conformational ensemble characteristic of a polymer chain in a poor solvent, where intra-chain interactions are more favourable than residue-solvent interactions. We note that, relative to the intrinsically disordered human proteome, A1-LCD$^*$ is remarkably compact, with only 5\% of IDRs estimated to have $\nu\leq0.475$ \cite{tesei2024conformational}. The predicted $R_g$ of $2.62\pm0.03$~nm is in good agreement with the experimental value reported by Martin et al. \cite{martin2021interplay} (Fig.~\ref{fig:2}C). To further illustrate the conformational ensemble, we calculate the intra-chain contact map and observe long-range interactions (Fig.~\ref{fig:2}B): Residue pairs separated by more than 20 positions along the sequence are in contact in up to 4\% of the sampled conformations.
 
\subsection{Single-chain MDP simulation} \label{sec:ex_single_MDP}
In this example, we perform a single-chain simulation of the MDP hnRNPA1$^*$ and characterise its conformational ensemble. In addition to the input for simulating an IDR detailed in the previous section, for MDPs we provide a PDB file as the starting structure. This structure can be obtained from experiments or, for example, from AlphaFold predictions \cite{jumper2021highly}. To restrain the folded domains and maintain their structure throughout the simulation, we use an elastic network model (ENM), whereas the IDRs of the MDP are modelled as flexible chains. To apply the ENM, we first identify the boundaries of each structured domain, for example by visual inspection of the three-dimensional protein structures. Each domain is then mapped to a sequence segment delimited by a start and an end residue. Only pairs of non-neighbouring residues within the same domain are restrained by the harmonic potential of the ENM. To provide the definitions of the domain boundaries, we create a domain file (\verb|domains.yaml| by default) containing the following lines:
\begin{verbatim}
hnRNPA1S:
 - [11,89]
 - [105,179]
\end{verbatim}
\verb|hnRNPA1S| is the protein name, and \verb|[11,89]| and \verb|[105,179]| are two restrained domains: The first spanning residues 11--89 and the second residues 105--179 (1-based indexing, inclusive of the termini).\footnote{The restraining algorithm assumes that residue indices in the PDB are numbered starting from 1. For consistency, users should consider renumbering \texttt{resSeq} in their PDB files starting from \texttt{1} for the first residue.}
For proteins that have long loops protruding from within a domain, one may exclude such loops in the domains by using nested lists:
\begin{verbatim}
SNAP_FUS:
 - [286, 368]
 - [423, 451]
 - [[537, 564], [586, 701]]
\end{verbatim}
In this case, the protein \verb|SNAP_FUS| has three restrained domains, of which the third contains a loop (residues 565--585) that we do not restrain.

We specify the MDP component as for simulations of single IDRs (Section~\ref{sec:ex_single_IDR}), with the following modifications:

\begin{verbatim}
components = Components(
    restraint=True,                     # apply restraints
    fresidues='{cwd}/input/residues_CALVADOS3.csv',  
    fdomains='{cwd}/input/domains.yaml',
    pdb_folder='{cwd}/input',                  
    use_com=True,                       # COM representation
    restraint_type='harmonic',           
    k_harmonic=700,                     # force constant
    cutoff_restr=0.9,
    )
\end{verbatim}
\verb|restraint=True| indicates that the protein will be regarded as an MDP. For simulations of systems containing MDPs, we recommend using \verb|residues_CALVADOS3.csv|, which was parametrised using experimental data for both IDRs and MDPs \cite{cao2024coarsegrained}. We then use \verb|fdomains| and \verb|pdb_folder| to specify the path of the file containing the domain boundaries and the folder containing the PDB files. The name of the PDB file in \verb|pdb_folder| should match the name of the component. With \verb|use_com=True|, we set the centre-of-mass representation for the restrained folded domains. We define the ENM with \verb|restraint_type='harmonic'|, set the force constant using \verb|k_harmonic| (default: \SI{700}{\kJ\per\mol}), and apply restraints to residue pairs separated by up to \verb|cutoff_restr| (default: \SI{0.9}{\nano \meter}).
As in the previous example, we performed three independent simulations and used the SD over the ensemble averages across the replicas as an estimate of the sampling error. To calculate per-frame $R_{ee}$ and $R_g$ values, and the average contact map, we included the same lines in \verb|prepare.py| as for the example in Section~\ref{sec:ex_single_IDR}, setting \verb|is_idr=False| to skip the calculation of $\nu$.

The $R_g$ distribution of full-length hnRNPA1$^*$ is shifted toward larger values compared to that of its C-terminal LCD (Fig.~\ref{fig:2}C) and has a mean of $R_g=3.27\pm\SI{0.01}{\nano\meter}$, in good agreement with the experimental measurement ($3.12\pm\SI{0.08}{\nano\meter}$) \cite{martin2021interplay,thomasen2022improving}. The intra-chain contact map highlights long-range interactions between the folded RRMs and the LCD (Fig. \ref{fig:2}D), consistent with previous observations \cite{martin2021interplay}.
A simulation of full-length hnRNPA1$^*$ performed with the settings detailed in this section has a speed of 250\,ns/day on a single core (750\,ns/day on 4 cores and 1000\,ns/day on 8 cores) of an AMD EPYC 7552 CPU and \SI{36.4}{\micro\second}/day on an NVIDIA A40 GPU. 

\subsection{Simulation of two (or more) different IDPs} \label{sec:ex_two_IDR}

CALVADOS can be used to simulate systems of several IDPs to predict the strength and mode of IDP-IDP interactions, as well as the influence of interaction partners on the conformational properties of individual chains. In the simplest case, we can simulate two chains with identical sequences and characterise the self-interaction of the IDP. Here, we simulate a heterogeneous two-chain system consisting of one copy of $\alpha$-synuclein ($\alpha$-Syn) and one copy of Tau35, a fragment that spans residues 187--441 of full-length human tau (ht2N4R). The steps outlined in the following can also be used to simulate and analyse more complex mixtures with several copies of many different IDPs. Two-chain simulations of $\alpha$-Syn (sequence length 140 residues) and Tau35 (sequence length 255 residues) run with a speed of \SI{30}{\micro\second}/day on an NVIDIA V100-16GB GPU with the settings described below.

\begin{figure}[tb]
\centering
\includegraphics[width=4.5in]{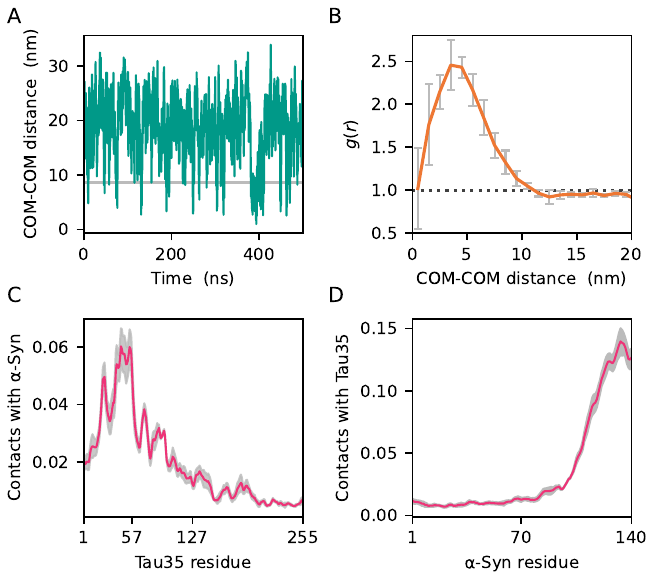}
\caption{Simulation of two different IDPs. (A)~Time series of aSyn-Tau35 COM--COM separation shown for \SI{500}{\nano\second} (5,000 simulation frames). Grey horizontal line: sum of the average radii of gyration of $\alpha$-Syn and Tau35. (B)~Radial distribution function, $g(r)$, as a function of the aSyn-Tau35 COM--COM separation, $r$. $g(r)$ is averaged across three independent simulation trajectories, and error bars show the SD across the three runs. Dotted black line: $g(r) = 1$. (C)~Total number of time-averaged contacts formed between residues in Tau35 and any residue in $\alpha$-Syn. (D)~Total number of time-averaged contacts formed between residues in $\alpha$-Syn and any residue in Tau35. The total contacts are averaged across three independent simulations, and grey shaded areas show the SD across the three replicas.}
\label{fig:3}
\end{figure}

In the \verb|prepare.py| script, we instantiate the \verb|Config| class as illustrated above for single-chain systems, for example with a box size and solution conditions to match experimental data. In this example, we used \verb|box=[40, 40, 40]|, \verb|wfreq=10000|, \verb|steps=5e8|, \verb|temp=288|, \verb|ionic=0.12|, and \verb|pH=7.2|. The system contains two components, one for each type of protein. We specify the component defaults as for the single-IDR example (Section~\ref{sec:ex_single_IDR}), and then add each component to our system with \verb|components.add()|:
\begin{verbatim}
components = Components(
    ... # defaults as for the single-chain IDR example       
    )
components.add(name='aSyn')     # add single copy of aSyn 
components.add(name='Tau35')    # add single copy of Tau35
\end{verbatim}
To characterise the interaction between the two IDPs, we used the two-chain simulation trajectory to calculate the radial distribution function, $g(r)$, and the residue-residue contacts between the proteins. $g(r)$ is a function of the inter-chain separation, $r$, which is commonly computed as the distance between the COMs of the two chains. We can add the following lines in \verb|prepare.py| to run the analysis functions \verb|calc_contact_map()| and \verb|calc_com_traj()| immediately after the simulation is completed to generate a trajectory file in which each IDP is represented by its COM, and to calculate the inter-chain contact map:
\begin{verbatim}
subprocess.run(f'mkdir -p aSyn_Tau35',shell=True)
subprocess.run(f'mkdir -p aSyn_Tau35/data',shell=True)
analyses = f"""
from calvados.analysis import calc_com_traj, calc_contact_map
# dictionary of chain indices
# 0-based indexing
chainid_dict = dict('aSyn' = 0, 'Tau35' = 1)
calc_com_traj(
    path='aSyn_Tau35', sysname='aSyn_Tau35', output_path='data',
    residues_file='{cwd}/input/residues_CALVADOS2.csv',
    chainid_dict=chainid_dict, start=10
    )
calc_contact_map(
    path='aSyn_Tau35', sysname='aSyn_Tau35',
    output_path='data', chainid_dict=chainid_dict
    )
"""
config.write(path,name='config.yaml',analyses=analyses)
\end{verbatim}
Using the COM-based trajectory, we compute COM--COM separations for each simulation frame (Fig.~\ref{fig:3}A) and use these to calculate $g(r)$ (Fig.~\ref{fig:3}B). The second virial coefficient, $B_{2}$, can be calculated from $g(r)$ through the following integral:
\begin{equation}
B_{2} = -2 \pi \int_{0}^{R_c} [g(r) - 1] r^{2}dr,  
\end{equation}
where $r$ is the COM--COM separation, and the upper limit of integration, $R_c$, is set to half the edge length of the cubic simulation box or less. For large values of $r$, $g(r)$ should approach one, as the interactions between the chains vanish. However, for systems of finite size, $g(r)$ can deviate significantly from one also at large values of $r$ (Fig.~\ref{fig:3}B). In our example, the attractive interaction between the two chains leads to a local accumulation at short separations, which in turn causes a depletion at larger separations, relative to the bulk concentration in the simulation box. Different methods have been proposed to correct for this finite-size effect prior to calculating $B_{2}$ \cite{ganguly_convergence_2013,heidari_refined_2024}. We here used the correction proposed by Ganguly and van der Vegt \citep{ganguly_convergence_2013} and obtained $B_2=-2\pm1$ L mol kg$^{-2}$.

While $B_{2}$ indicates the strength of the inter-chain interaction, and whether this is net attractive ($B_{2}<0$) or net repulsive ($B_{2}>0$), our simulations also provide information on which residues are involved in the formation of the transient dimer. By summing the two-dimensional contact map along each axis, we calculate the time-averaged contacts formed by each residue of one chain with all the residues of the other (Fig.~\ref{fig:3}C and D). Our simulations reveal that the residues engaging in transient inter-chain interactions are predominantly in the N-terminal region of Tau35 and in the C-terminal region of $\alpha$-Syn. This is in accordance with experimental NMR chemical shift perturbations, which show that $\alpha$-Syn and full-length human tau interact via the negatively charged C-terminal domain of $\alpha$-Syn and the positively charged P2 region of full-length tau (residues 12--57 in Tau35) \cite{siegert_interplay_2021}. 

\subsection{Single-component slab simulation} \label{sec:ex_slab_IDR}

In this example, we show how to simulate and analyse a system of multiple IDR chains that phase separate and form a protein-rich condensed (dense) phase in equilibrium with a dilute phase.
To reduce the impact of finite-size effects when simulating the coexistence of these two phases, we insert the proteins into an elongated box with equal side lengths along the $x$- and $y$-axes ($L_x=L_y$), and a length along the $z$-axis ten times larger ($L_z=10 \times L_x$). With such a simulation cell, proteins may spontaneously phase separate into a single slab spanning the periodic images along the $x$- and $y$-axes \citep{blas2008vapor,dignon2018sequence}. This setup minimises the interface area and allows us to approximate a bulk dense phase with only one or a few hundred copies of a protein. The long $L_z$ is needed to sample the dilute phase which has a concentration much smaller than the dense phase.

The key setting for a slab simulation are shown in the following snippet: 
\begin{verbatim}
config = Config(
    ... # define temp, ionic, pH
    box=[15, 15, 150],    # nm     
    topol='slab',         # place chains in a slab
    slab_width=20,        # of width 20 nm
    slab_eq=True,         # apply linear potential
    steps_eq=5000000,     # for this many steps
    wfreq=50000,
    steps=5400*50000      # 2.7 µs
    )
\end{verbatim}
With \verb|topol='slab'|, we speed up the equilibration of the two-phase system by first inserting the chains between $z=-10$ nm and $z=10$ nm (\verb|slab_width=20|), with $x$- and $y$-coordinates on a regular grid. We then apply a linear external potential $u_\text{ext}(z) = k\, |z-L_z/2|$, with $k_\text{eq}=0.02$ kJ mol$^{-1}$ nm$^{-1}$, to focus the chains towards the midplane of the box ($z=0$). The potential is applied for the first \SI{50}{\nano\second} (\verb|steps_eq=5000000|), during which the trajectory is saved to a file with prefix `equilibration\_'. Subsequently, we remove the restraint and simulate the systems for additional \SI{2.7}{\micro\second}, of which the initial \SI{0.2}{\micro\second} are discarded for equilibration. Simulations of A1-LCD, performed with the settings detailed in this section, have a speed of \SI{5.8}{\micro\second}/day on an NVIDIA A40 GPU.

\begin{figure}[tb]
\centering
\includegraphics[width=4.5in]{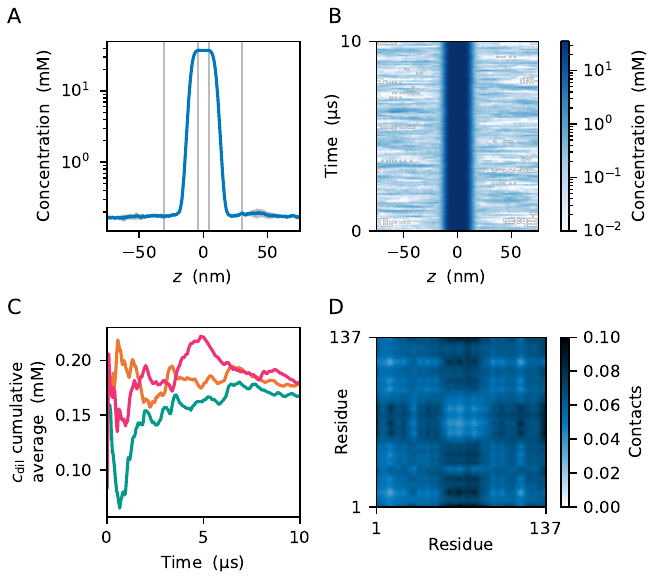}
\caption{Single-component slab simulation. (A)~Concentration profile of hnRNPA1 LCD along the $z$-axis of the simulation cell. The profile is averaged over three simulation replicas with the grey shaded areas showing the SD. Grey vertical lines: boundaries defining the regions within which we calculate the concentration in the dense phase, $c_\text{den}$, and in the dilute phase, $c_\text{dil}$. (B)~Instantaneous concentration profiles as a function of simulation time. (C)~Cumulative averages of $c_\text{dil}$ from three independent simulation replicas. (D)~Contact map between the chain closest to the mid-plane of the protein-dense slab and the surrounding chains.}
\label{fig:4}
\end{figure}

In the package, we implemented the class \verb|SlabAnalysis| which features basic routines to analyse slab simulations, as shown in the snippet below.
\begin{verbatim}
from calvados.analysis import SlabAnalysis
slab = SlabAnalysis(
            name='A1LCD', input_path='A1LCD',
            output_path='data', ref_name='A1LCD',
            verbose=True
            )
slab.center(start=400, center_target='all')
slab.calc_profiles()
slab.calc_concentrations(pden=2, pdil=8)
\end{verbatim}
First, we run \verb|center()| to calculate instantaneous concentration profiles in the production run (\verb|start=400|) and use these to shift the positions of all the beads, so as to align the slab in the middle of the box in each frame. With the resulting centred trajectory, we run \verb|calc_profiles()| to recalculate the instantaneous concentration profiles and their average over time (Fig.~\ref{fig:4}A and B). When the simulation converges, the profile is approximately symmetric around $z=0$. To obtain the concentrations in the coexisting phases, we run \verb|calc_concentrations()|. This function finds the position of the dividing surface, $z_\text{DS}$, and the thickness of the interface, $2\cdot d$, by fitting the half profiles for $z<0$ and $z>0$ to the sigmoidal function
\begin{equation}
c(z) = \frac{\widehat{c_\text{den}}+ \widehat{c_\text{dil}}}{2}+\frac{ \widehat{c_\text{dil}}-\widehat{c_\text{den}}}{2}\tanh{\left ( \frac{|z|-z_\text{DS}}{d} \right )},
\end{equation}
where $\widehat{c_\text{den}}$ and $\widehat{c_\text{dil}}$ are estimates of the average concentrations of the dense and dilute phases, respectively.
Using the best-fit parameters, \verb|calc_concentrations()| then calculates the time series of $c_\text{den}$ and $c_\text{dil}$ from the mean concentrations in $|z|< z_\text{DS}-p_\text{den}\cdot d$ and $|z|>z_\text{DS}+p_\text{dil}\cdot d$, where $p_\text{den}$ and $p_\text{dil}$ can be defined by the user and are set by default to 2 and 8, respectively. The output generated by these functions includes the slab-centred trajectory, the time-averaged concentration profile (Fig.~\ref{fig:4}A), the time series of the concentration profile (Fig.~\ref{fig:4}B), the time series of $c_\text{den}$ and $c_\text{dil}$, and a table summarizing the settings and results of these analyses. 

The saturation concentration, $c_\text{sat}$, of the dilute phase at equilibrium with a biomolecular condensate is particularly sensitive to changes in amino acid sequence and solution conditions \cite{pappu2023phase}, and is often used to quantify the propensity of a biomolecule to phase separate. Therefore, $c_\text{sat}$ is a key parameter to benchmark simulations against experiments. In this example, we simulated the 137-residue-long LCD of hnRNPA1 (A1-LCD), without the nuclear localization signal \citep{martin2020valence,bremer2022deciphering}, in three independent simulation replicas, setting \verb|temp=293| and \verb|ionic=0.15|. In good agreement with the experimental $c_\text{sat}$ value of $102.2\pm0.4\,\si{\micro\molar}$ \cite{martin2020valence,bremer2022deciphering}, we estimate $c_\text{dil}=176\pm6\,\si{\micro\molar}$ as the mean $\pm$ SD across the three replicas. To ensure the convergence of the mean $c_\text{dil}$, a simulation time of around \SI{7}{\micro\second} is required for this system, as shown by the cumulative averages of $c_\text{dil}$ for the three replicas (Fig.~\ref{fig:4}C).

In addition to estimating thermodynamic properties, such as $c_\text{sat}$ and transfer free energy, $\Delta G_\text{trans}=RT\ln{(c_\text{dil}/c_\text{den})}$, slab simulations provide molecular-level insight into intermolecular interactions in the coexisting phases. Using the function \verb|calc_com_traj()| introduced in Section~\ref{sec:ex_slab_IDR}, we obtain the trajectory of the centres of mass of all the chains in the system. The function \verb|calc_contact_map()| uses this file to find, in each frame, the chain that is closest to the midplane of the slab, and calculates the residue-residue distances, $r_{ij}$, between this chain and all surrounding chains. We convert these distances into contacts using the switching function in Eq.~\ref{eq:switch}.
The contact map for A1-LCD in the condensate (Fig.~\ref{fig:4}D) highlights the strong attractive interactions between the positively charged N-terminal and C-terminal regions and the negatively charged and aromatic-rich region between residues 50 and 90.

\begin{verbatim}
# calculate homotypic contact map
from calvados.analysis import calc_com_traj, calc_contact_map
# dictionary of chain indices
# 0-based indexing, inclusive
chainid_dict = dict(A1LCD = (0, 99))
calc_com_traj(
    path=f'{cwd}/A1LCD', sysname='A1LCD',
    output_path=f'{cwd}/data',
    residues_file=f'{cwd}/input/residues_CALVADOS2.csv',
    chainid_dict=chainid_dict
    )
calc_contact_map(
    path=f'{cwd}/A1LCD', sysname='A1LCD',
    output_path=f'{cwd}/data',
    chainid_dict=chainid_dict,
    is_slab=True
    )
\end{verbatim}

% \clearpage
\subsection{Slab simulation of mixed systems} \label{sec:ex_slab_mixed}
Many IDRs undergo phase separation with nucleic acids; in this section, we simulate an example of such a mixed system, consisting of the RGG3 domain of Fused in Sarcoma (FUS-RGG3) and a 40-base polyuracil strand (polyU40). Although FUS-RGG3, which is positively charged, does not easily phase separate alone, the addition of moderate amounts of RNA induces phase separation \cite{kaur2021sequence}. Since the two-bead-per-residue RNA model is sequence independent, a specific strand is defined solely by its length; in this case, by specifying 40 consecutive \verb|r| characters in the FASTA file (\verb|idr.fasta|). In the script \verb|prepare.py|, we set up a system containing 200 chains of FUS-RGG3 and 60 chains of polyU40 and describe these using the CALVADOS~2 model for proteins \cite{tesei2023improved} and the CALVADOS-RNA model \cite{yasuda2024coarsegrained} via \verb|residues_C2RNA.csv|. The instantiation of \verb|Components| and the lines for adding the chains are as follows:
\begin{verbatim}
components = Components(
    fresidues=f'{cwd}/input/residues_C2RNA.csv',
    ffasta=f'{cwd}/input/idr.fasta',
    rna_kb1=1400.0, rna_kb2=2200.0,
    rna_ka=4.20, rna_pa=3.14,
    rna_nb_sigma=0.4, rna_nb_scale=15,
    rna_nb_cutoff=2.0
    )
components.add(
    name='FUS-RGG3', molecule_type='protein', 
    nmol=100,  charge_termini='both'
    ) 
components.add(
    name='polyU40', molecule_type='rna', nmol=60
    )
\end{verbatim}
The parameters set through the \verb|Config| object are the same as for the single-component system of Section~\ref{sec:ex_single_IDR}, except for \verb|box=[15, 15, 80]|, \verb|wfreq=100000|, \verb|steps=100000000|, and the solution conditions, which we match to those of Kaur et al.~\cite{kaur2021sequence}, namely \verb|temp=293.15|, \verb|ionic=0.15|, and \verb|pH=7.5|.
% \begin{verbatim}
% config = Config(
%   # GENERAL
%   sysname = sysname, # name of simulation system
%   box = [Lx, Lx, Lz], # nm
%   temp = 293.15, # 20 degrees Celsius
%   ionic = 0.15, # molar
%   pH = 7.5, # 7.5
%   topol = 'random',
%   slab_eq = True,
%   k_eq = 0.02,
%   steps_eq = 1000000,
% \end{verbatim}
The performance of this simulation is \SI{7.5}{\micro\second}/day using an NVIDIA GeForce RTX 3090 GPU and a single thread (AMD Ryzen Threadripper 3960X 24-Core Processor). 

\begin{figure}[tb]
    \centering
    \includegraphics[width=4.5in]{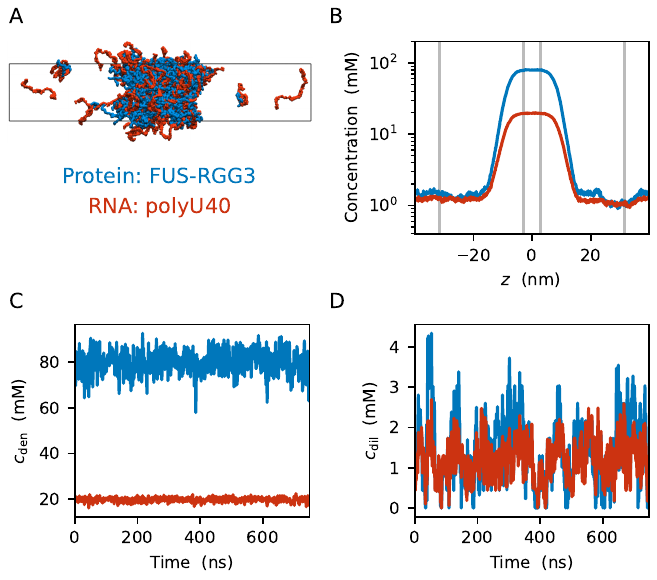}
    \caption{Slab simulation of protein and RNA. (A)~A representative snapshot showing FUS-RGG3 in blue and polyU40 in red. (B)~Concentration profile of FUS-RGG3 and polyU40 along the $z$-axis. Grey vertical lines: boundaries of dilute and dense phases. Time series of the concentration of FUS-RGG3 and polyU40 in the (C)~dense and (D)~dilute phases.}
    \label{fig:5}
\end{figure}

The simulation trajectory can be analysed to determine the distributions of proteins and RNA chains in the dilute and protein-dense phases. As in the previous example, \verb|center()| removes the overall motion of the condensate by translating its centre to the midplane of the elongated box (Fig.~\ref{fig:5}A). After discarding the initial 250 steps for equilibration, \verb|calc_profiles()| reads the translated trajectory and computes the concentration profiles of FUS-RGG3 and polyU40 (Fig.~\ref{fig:5}B). 
Finally, \verb|calc_concentrations()| identifies the dilute and dense phase regions and computes their mean concentration for each frame (Figs.~\ref{fig:5}C and D). From the density profiles and concentration time series (Figs.~\ref{fig:5}B and C), we estimate the concentration ratio of FUS-RGG3 to RNA within the condensate to be approximately 4:1. 
These analyses can be executed by adding the following lines to \verb|prepare.py|:
\begin{verbatim}
from calvados.analysis import SlabAnalysis
slab_analysis = SlabAnalysis(
    name='mixed_system',
    input_path=f'{cwd}/mixed_system', 
    output_path=f'{cwd}/data',
    input_pdb='top.pdb', input_dcd=None,
    centered_dcd='traj.dcd',
    # use proteins as reference for centering
    ref_chains=(0, 199),  # 0-based indexing, inclusive
    ref_name='FUS-RGG3',
    client_chain_list=[(200, 259)],  
    client_names=['polyU40'],
    verbose=False
    )
slab_analysis.center(
    start=250,
    center_target='ref'
    )     
slab_analysis.calc_profiles()
slab_analysis.calc_concentrations() 
\end{verbatim}

\subsection{Simulations with crowders} \label{sec:ex_slab_crowder}

\begin{figure}[tb]
\centering
\includegraphics[width=4.5in]{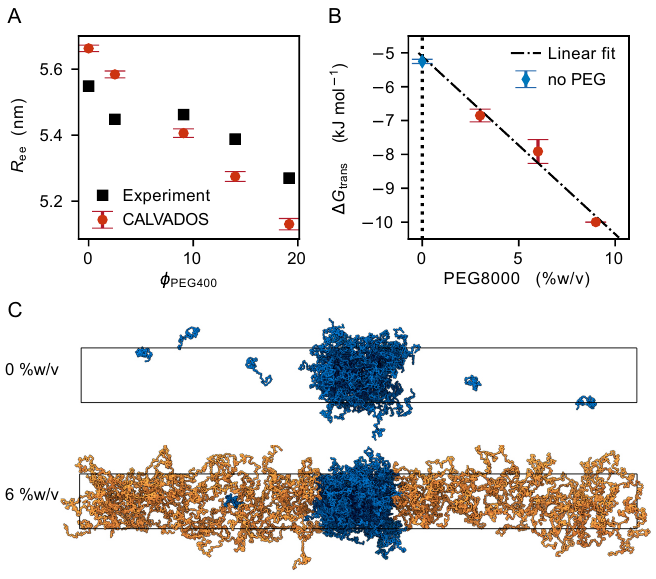}
\caption{Simulations of IDPs in the presence of PEG. (A)~End-to-end distances ($R_\text{ee}$) of a single chain of ACTR at increasing concentrations of PEG400. The figure shows previously generated simulation data \citep{rauh2025crowder} and a comparison to experimental data \cite{soranno_single-molecule_2014}.
(B)~Phase separation propensity ($\Delta G_\text{trans}$) of A1-LCD with PEG8000 (red dots). A linear fit to the data generated at three different concentrations of PEG is used to extrapolate the phase separation propensity in the absence of PEG ($y$-intercept = $-5.1\pm0.6 \text{kJ~mol}^{-1}$). Blue diamond: $\Delta G_\text{trans}$ from a simulation of A1-LCD in the absence of PEG ($-5.25\pm0.06 \text{kJ~mol}^{-1}$).
(C) Representative snapshots of slab simulations of A1-LCD without PEG8000 (100:0) and with 6~\%w/v PEG8000 (100:152).}
\label{fig:6}
\end{figure}

The environment of an IDP can have a significant effect on its conformational dynamics. To study the effects of macromolecular crowding in the CALVADOS model, we have implemented a model for the synthetic crowding agent polyethylene glycol (PEG) \citep{rauh2025crowder}. In our PEG model each bead represents a single residue, and PEG polymers of different molecular weights are represented by sequences of varying lengths, with the letter \verb|J| denoting the PEG monomer in the FASTA file.
To illustrate what effects can be modelled, we here simulate two different systems, (i)~a single chain of an IDR (ACTR) with increasing concentrations of PEG400, and (ii)~a slab of A1-LCD with increasing concentrations of PEG8000.
In the first case, our model is applied to study how crowding affects the global dimensions of an IDP (Fig.~\ref{fig:6}A). In the second case, we show how a PEG-titration can be used to determine the phase separation propensity of an IDP in the absence of PEG (Fig.~\ref{fig:6}B).

Before starting the simulations, we first calculate the number of PEG chains, $N_\text{PEG}$, as
\begin{equation}
\label{eq:n_chains}
N_\text{PEG}(\%\text{w/v PEG}) = \frac{10\times\text{\%w/v}}{\text{MW}_\text{PEG}}\times V_\text{box}\times N_{A},
\end{equation}
where $\%w/v$ is the PEG mass concentration in g/100~mL, $\text{MW}_\text{PEG}$ is the molecular weight of PEG, $V_\text{box}$ is the volume of the simulation box, and $N_\text{A}$ is Avogadro's number. For direct comparison with experiments, we convert $\%w/v$ into the volume fraction, $\phi_\text{PEG}$, using $\phi_\text{PEG}=(10\times\text{\%w/v})/\rho_\text{PEG,bulk}$, where $\rho_\text{PEG,bulk}=1,120$~g/L \citep{soranno_single-molecule_2014,rauh2025crowder}.

To set up the simulations, the configuration settings are defined as for a standard slab simulation except for the PEG-model specific parameter \verb|fixed_lambda|, which fixes the AH stickiness parameter $\lambda$ to 0.2 for PEG--PEG and protein--PEG interactions \citep{rauh2025crowder}. 
\begin{verbatim}
config = Config(
    ... # general definitions
    topol='grid',
    fixed_lambda=0.2
    )
\end{verbatim}

Additionally, the bead size $\sigma$ and the MW of a PEG monomer are defined in \verb|residues_C2PEG.csv| and read as usual:
\begin{verbatim}
components = Components(
    ...
    fresidues=f'{cwd}/input/residues_C2PEG.csv'
)
\end{verbatim}

To simulate ACTR with PEG, we use the experimental conditions \citep{soranno_single-molecule_2014}, namely \verb|temp=295.15|, \verb|ionic=0.11| and \verb|pH=7.5|. 
We add a single chain of ACTR (\verb|nmol=1|) and $N_\text{PEG}$ molecules of PEG400 (\verb|nmol=N_PEG|) as follows: 
\begin{verbatim}
components.add(name='ACTR', molecule_type='protein', nmol=1)
components.add(
    name='PEG400', molecule_type='crowder',
    nmol=N_PEG, charge_termini=False
    )
\end{verbatim}

The conformational properties of the IDR are analysed as described in Section~\ref{sec:ex_single_IDR}, now selecting only the protein chain in \verb|save_conf_prop()|, by specifying \verb|select='not resname PEG'|. 
The end-to-end distance of ACTR, $R_\text{ee}$, as a function of $\phi_\text{PEG400}$ shows a decrease of $R_\text{ee}$ induced by the crowder (Fig.~\ref{fig:6}A).

With minor changes to the input shown in Section~\ref{sec:ex_slab_IDR}, we can perform slab simulations with PEG to mimic a PEG titration experiment and determine, by extrapolation, the phase separation propensity ($\Delta G_\text{trans}$) of a protein that is weakly prone to phase separate. Instead, in this example, we simulate the same IDR as in Section~\ref{sec:ex_slab_IDR} and compare the value extrapolated from the PEG titration to $\Delta G_\text{trans}$ calculated from slab simulations for the IDR without any PEG.

As for the single-component system, the topology keyword is set to \verb|slab| and we specify \verb|slab_width=20| to insert the proteins between $z=-10$~nm and $z=10$~nm. In addition, \verb|slab_outer=25| places the crowder molecules around the slab at $\vert z \vert > 25$~nm.

\begin{verbatim}
config = Config(
    ... # general definitions as above
    box=[15, 15, 150],  # nm
    topol='slab',       # place proteins in a slab
    slab_width=20,      # of width 20 nm
    slab_outer=25,      # and the crowder outside
    fixed_lambda=0.2,
)
...
components.add(name='A1LCD', molecule_type='protein', nmol=100)
components.add(
    name=f'PEG8000', molecule_type='crowder',
    nmol=N_PEG, charge_termini=False
    )
\end{verbatim}

The performance of these protein--crowder simulations ranges from \SI{4.75}{\micro\second}/day without PEG to \SI{3.8}{\micro\second}/day with 3~\%w/v PEG8000 to \SI{2.1}{\micro\second}/day at 9~\%w/v PEG8000 on a NVIDIA Tesla V100. % PCIe 16GB.  
The density profiles and concentrations of protein and PEG in the dense and dilute phases are calculated as for the mixed protein--RNA simulation in Section~\ref{sec:ex_slab_mixed}.
\begin{verbatim}
from calvados.analysis import SlabAnalysis
slab = SlabAnalysis(
    name='A1LCD_PEG8000', 
    input_path=f'{cwd}/A1LCD_PEG8000',
    output_path=f'{cwd}/data',
    ref_name='A1', ref_chains=(0, 99),
    client_names=['PEG8000'], 
    client_chain_list=[(100, 99 + N_PEG)]
    )
slab.center(
    center_target='ref' # for centering only on A1LCD
    )
slab.calc_profiles()
slab.calc_concentrations()
\end{verbatim}

To determine the phase-separation propensity of the protein in the absence of PEG, we perform a linear fit and extrapolate $\Delta G_\text{trans}$ without PEG as the $y$-intercept. For the case of A1-LCD, we find that this approach reproduces the value from a simulation without PEG (Fig.~\ref{fig:6}B).

\subsection{Custom restraints} \label{sec:ex_custom_restraints}

\begin{figure}[tb]
\centering
\includegraphics[width=4.5in]{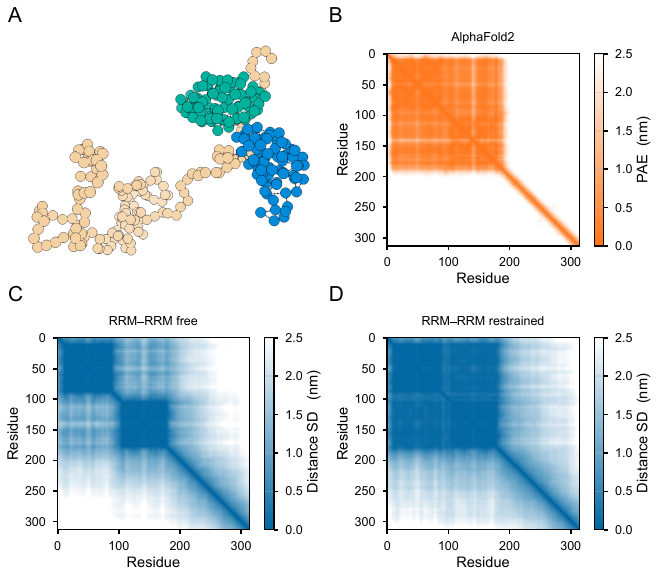}
\caption{Full-length hnRNPA1$^*$ simulations without and with custom restraints between key residues at the RRM–RRM interface. (A)~Snapshot of coarse-grained full-length hnRNPA1$^*$ with the RRMs coloured in green and blue. (B)~AlphaFold2 predicted aligned error (PAE). SD of residue–residue distances (C)~without or (D)~with RRM-RRM restraints.}
\label{fig:7}
\end{figure}

The CALVADOS package allows users to define custom restraints between any pairs of residues in the system. In \verb|prepare.py|, custom restraints are enabled via
\begin{verbatim}
config = Config(
    ... # general definitions
    custom_restraints=True,
    custom_restraint_type='harmonic',
    fcustom_restraints=f'{cwd}/input/cres.txt',
)
\end{verbatim}
where the text file \verb|cres.txt| contains the list of custom restraints. As an example, we restrain the two RRM domains in full-length hnRNPA1$^*$ \citep{martin2021interplay} to move as a single domain \citep{ritsch2021characterization} and analyse the inter-domain fluctuations via the SD of the residue pair distances across the simulation. The \verb|cres.txt| file reads
\begin{verbatim}
hnRNPA1S 1 72 | hnRNPA1S 1 157 | 0.559 700.0
hnRNPA1S 1 73 | hnRNPA1S 1 161 | 1.072 700.0
hnRNPA1S 1 75 | hnRNPA1S 1 155 | 0.765 700.0
...
\end{verbatim}
with the syntax 
\begin{verbatim}component1 copy1 residue1 | component2 copy2 residue2 | r0 k
\end{verbatim}
defining a harmonic bond with parameters as in Eq.~\ref{eq:2}. In this example, we only have one copy of a single component and restrain residues 72 with 157, 73 with 161, etc. using the corresponding COM-COM separations in the input structure (in nm) as the equilibrium distances. These restraints strongly affect how the RRMs move together in the simulation (Fig.~\ref{fig:7}). Whereas the unrestrained RRMs tumble more independently than suggested before \citep{martin2021interplay,ritsch2021characterization} and by the AlphaFold2 predicted aligned error (PAE) matrix (Fig.~\ref{fig:7}B~and~C), the custom restraints cause the RRMs to move together (Fig.~\ref{fig:7}D). 
\footnote{We recommend using domain definitions (\texttt{domains.yaml}) rather than custom restraints (\texttt{cres.txt}) to restrain the residues within entire folded domains (see Section~\ref{sec:ex_single_MDP}) and to reserve the use of custom restraints to circumvent complicated domain definitions when only a few restraints are needed between the same or different molecules.}

\section{Modification and extension of the package} \label{sec:extension}

\begin{figure}[tb]
\centering
\includegraphics[width=4.5in]{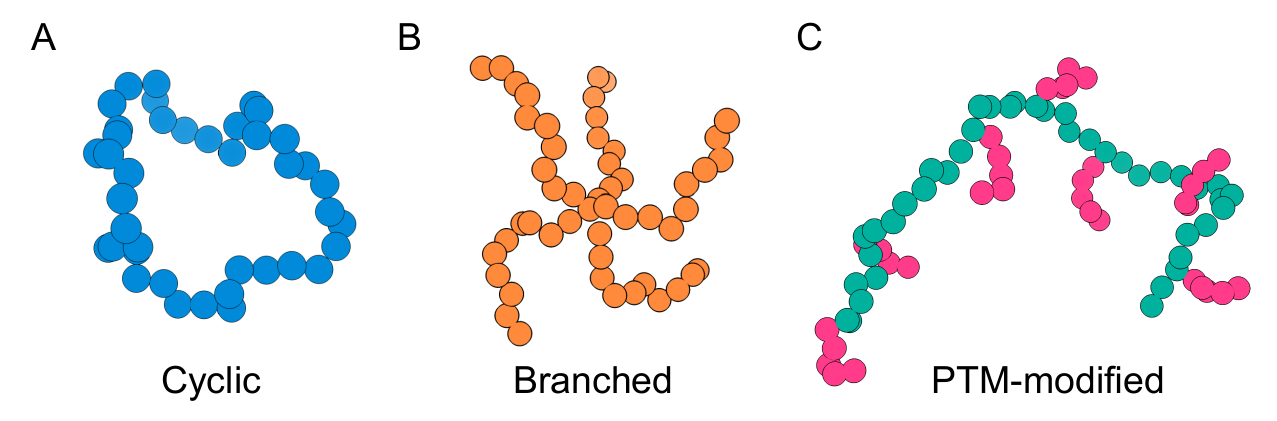}
\caption{Example of how CALVADOS may be extended. Each of the examples is a subclass of \texttt{Protein} with minor modifications to bond definitions (A)~cyclic, (B)~branched, and (C)~PTM-modified.}
\label{fig:8}
\end{figure}

The modular architecture of the software allows for modifications and extensions, and contributions from the community are welcome. The CALVADOS code is hosted on GitHub (\verb|github.com/KULL-Centre/CALVADOS|) and is made available with a GNU General Public License v3.0. We show three examples of simple CALVADOS extensions: Cyclic peptides, star-shaped polymers and residues with post-translational modifications (Fig.~\ref{fig:8}).

% \subsection{Cyclic peptides}
As the first example, we derive a new component class \verb|Cyclic| that inherits from the \verb|Protein| component class. Here, we change the criteria for the bond definitions by modifying \verb|bond_check()| from
\begin{verbatim}
condition = (j == i + 1)
\end{verbatim}
to
\begin{verbatim}
condition0 = (j == i + 1)
condition1 = (j == self.nbeads - 1) and (i == 0)
condition = condition0 or condition1
\end{verbatim}

In this way, a bond is defined between the first and last residue (\verb|self.nbeads - 1|) as well as between neighbouring residues in the sequence, creating a cyclic peptide (Fig.~\ref{fig:8}A). 

As the second example, we create branched star-shaped polymers, again by modifying the \verb|bond_check()| method. We define a central bead, adding bonds to that bead and removing bonds at the end of the `arms' (Fig.~\ref{fig:8}B).
\begin{verbatim}
if self.n_ends in [0, 1, 2]:
    return super().bond_check(i, j)
else:
    if (self.nbeads - 1) % self.n_ends == 0:
        branch_length = int((self.nbeads - 1) / self.n_ends)
    else:
        branch_length = int((self.nbeads - 1) / self.n_ends) + 1
    condition0 = (j == i + 1) and ((j - 1) % branch_length != 0)
    condition1 = (i == 0) and ((j - 1) % branch_length == 0)
    condition = condition0 or condition1
    return condition
\end{verbatim}
The number of `arms' of the branched `\verb|seastar|' polymer can be chosen as an attribute \verb|n_ends| in the component definition in \verb|prepare.py|.
\begin{verbatim}
components.add(
    name='branched_polymer', 
    molecule_type='seastar', n_ends=5
    )
\end{verbatim}

As a final example, we introduce post-translational modifications (PTMs). There are at least two ways of incorporating PTMs into CALVADOS: As a first option, we modify or add amino-acid residue definitions to account for changes in charge or stickiness (e.g., caused by phosphorylation) without explicitly changing the number of beads. For CALVADOS, we parametrised phosphorylated serine (pSer) and threonine (pThr) beads with increased size and molecular weight, decreased stickiness, and partial charges computed based on solution pH and experimental p$K_a$ \cite{Hendus-Altenburger2019-gw,rauh2025phospho}. We determined stickiness parameters for pSer and pThr using a top-down approach, targeting experiments on global dimensions for a set of phosphorylated and unphosphorylated IDRs \citep{rauh2025phospho}. This procedure resulted in a model with $\lambda_\text{pSer}\approx0.09$ and $\lambda_\text{pThr}\approx0.00$ \citep{rauh2025phospho}, though we note that it may also be possible to generate parameters using bottom-up procedures \citep{perdikari2021predictive,lohberger2025hydrodynamic}.

Here, we show how to simulate the 10-fold phosphorylated IDR Ash1 with our phosphorylation model \citep{rauh2025phospho}.
The first step is to prepare the sequence in FASTA format with the single letter codes \verb|B| and \verb|O| for pSer and pThr, respectively. 
\begin{verbatim}
>10pAsh1
SASSBPBPSOPTKSGKMRSRSSBPVRPKAYOPBPRBPNYHR
FALDBPPQBPRRSSNSSITKKGSRRSSGSBPTRHTTRVCV
\end{verbatim}

The \verb|B| and \verb|O| residues are added to the residue definitions in \verb|residues_pCALVADOS2.csv|.
\begin{verbatim}
one, three,     MW, lambdas, sigmas,        q, bondlength
             ...                    ...
  B,   SEP, 165.04,  0.0925,  0.601,  -1.9686,       0.38
  O,   TPO, 179.07,  0.0013,  0.635,  -1.9406,       0.38
\end{verbatim}

Before starting the simulation, we compute the partial charges on pSer and pThr, using the Henderson-Hasselbalch equation, and overwrite \verb|residues_pCALVADOS2.csv| to update the \verb|q| values.
\begin{verbatim}
# set charge on pSer and pThr based on input pH
pKa_dict = dict(SEP = 6.01, TPO = 6.3)
df_residues = pd.read_csv(residues_file, index_col='three')
for pres in pKa_dict.keys():
    df_residues.loc[pres,'q'] = - 1 - 1 / (1 + 10**(pKa_dict[pres] - pH))
df_residues.reset_index().set_index('one').to_csv(residues_file)
\end{verbatim}

To analyse the global dimensions of the phosphorylated IDR, we use the \verb|save_conf_prop()| function, as described in Section~\ref{sec:ex_single_IDR}. To explore how the global dimensions change upon phosphorylation, we can simulate the unphosphorylated IDR, and calculate $\Delta R_\text{g}$ or $\Delta \nu$.   

As a second way of incorporating PTMs into CALVADOS (which can be combined with the first), larger PTMs such as ubiquitination, sumoylation, glycosylation, etc. could be introduced by adding additional branching points at specific residues on the protein chain (Fig.~\ref{fig:8}C).  As an example, we describe a simple case of adding linear PTMs to specific protein residues, but more complicated cases (e.g. branched PTMs for glycosylation, or dyes for FRET experiments \citep{holla2024identifying}) can be implemented.

In the example below, the user defines PEGylated proteins in the component definition in \verb|prepare.py|:
\begin{verbatim}
components.add(
    name='A1-PEGylated', 
    molecule_type='ptm_protein',
    ptm_name='PEG',
    ptm_locations=[5, 10, 15] # 1-based
    )
\end{verbatim}
The code searches for the entry with name \verb|ptm_name| in the same FASTA file that also contains the protein sequence. The positions of the protein residues to be PEGylated are specified in the list \verb|ptm_locations|.

Under the hood, the PTM-modified proteins are defined as a subclass \verb|PTMProtein| which again inherits from \verb|Protein|. \verb|PTMProtein| has modified versions of methods \verb|calc_comp_seq()| and \verb|bond_check()| to concatenate protein and PTM sequences and to account for the additional protein-PTM and PTM-PTM bonds, respectively.
% \begin{verbatim}
% def calc_comp_seq(self)
%     ...
%     self.ptm_seq = str(records[self.ptm_name].seq)
%     for ptm_idx in self.ptm_locations:
%         self.seq = self.seq + self.ptm_seq
%     ...

% def bond_check(self, i: int, j: int):
%     ...
%     ptm_seqlocs = []
%     # residue-PTM bond
%     for idx, ptm_loc in enumerate(self.ptm_locations):
%         ptm_seqloc = self.nbeads_protein + idx * len(self.ptm_seq) 
%         ptm_seqlocs.append(ptm_seqloc)
%         if (i == ptm_loc - 1) and (j == ptm_seqloc):
%             return True
    
%     # PTM-PTM bond
%     if i >= self.nbeads_protein:
%         if (j == i + 1) and (j not in ptm_seqlocs):
%             return True
%     return False
% \end{verbatim}

We provide the options \verb|cyclic|, \verb|seastar|, and \verb|ptm_protein| as options for \verb|molecule_type| in \verb|prepare.py| as starting points for possible modifications to CALVADOS. We note, however, that these three molecule types have not been parametrised or tested against experimental or simulation data.

\section{Conclusion}
The CALVADOS software package can be used for simulations of IDRs, MDPs, solutions crowded by PEG, RNA, and mixtures of all of the above. While we have here provided examples of how to run such simulations, we remind users to keep the limitations of the models in mind, and to explore the literature to determine the ranges of applicability of these and related models. In the future, we envision to increase the complexity of the possible simulated systems by parametrising new biomolecules via combinations of top-down and bottom-up approaches, keeping in mind that data scarcity and force field limitations make it difficult to parametrise all cross-interactions accurately. We therefore encourage users with concrete scientific simulation problems to adapt the force fields to the problem at hand, for example by applying custom restraints using information from experiments, or by introducing custom residues. 

\section*{Acknowledgements}
We thank Matteo Cagiada, Daniela de Freitas, Hamidreza Ghafouri, Isabel Gm{\"u}r, Yukti Khanna, Ana Panteli{\'c}, Carlos Pintado-Grima, Eva Smorodina, and Lukas Stelzl for testing the package and providing valuable feedback.
We acknowledge access to computational resources via a grant from the Carlsberg Foundation (CF21-0392), from the ROBUST Resource for Biomolecular Simulations (supported by the Novo Nordisk Foundation grant no. NF18OC0032608), and at the Biocomputing Core Facility at the Department of Biology, University of Copenhagen. S.v.B. acknowledges support by the European Molecular Biology Organisation through Postdoctoral Fellowship grant ALTF 810-2022. K.L.-L. acknowledges support from the Novo Nordisk Foundation via the Protein Interactions and Stability in Medicine and Genomics (PRISM) centre (NNF18OC0033950). The original development of CALVADOS was supported by the BRAINSTRUC structural biology initiative funded by the Lundbeck Foundation (R155-2015-2666).

% \backmatter
% \bmhead{Supplementary information}

% \bmhead{Acknowledgements}

\section*{Declarations}

K.L.-L. holds stock options in and is a consultant for Peptone. The remaining authors declare no competing interests.

% Some journals require declarations to be submitted in a standardised format. Please check the Instructions for Authors of the journal to which you are submitting to see if you need to complete this section. If yes, your manuscript must contain the following sections under the heading `Declarations':

% \begin{itemize}
% \item Funding
% \item Conflict of interest/Competing interests (check journal-specific guidelines for which heading to use)
% \item Ethics approval and consent to participate
% \item Consent for publication
% \item Data availability 
% \item Materials availability
% \item Code availability 
% \item Author contribution
% \end{itemize}
\linespread{1}\selectfont
\bibliography{references}

\end{document}